\documentclass[aps,prl,reprint,groupedaddress,10pt,twocolumn]{revtex4}
\usepackage{graphicx}
\usepackage{color}

\usepackage{bm}

\newcommand{\vex}[1]{\bm{\mathrm{#1}}}
\newcommand{\msf}[1]{\mathsf{#1}}
\newcommand{\Rhot}{\msf{R}}
\newcommand{\wRhot}{\widetilde{\mathsf{R}}}
\newcommand{\rr}{\mathfrak{r}}
\newcommand{\ii}{\mathfrak{i}}

\newcommand{\Dqcp}{\Delta_{\mathsf{QCP}}}
\newcommand{\tpb}{t_{\mathsf{pb}}}
\newcommand{\tl}{t_{\mathsf{3}}}
\newcommand{\e}{\varepsilon}
\newcommand{\Emin}{E_{\mathsf{min}}}
\newcommand{\G}{\mathcal{G}}

\def\8{\infty}

\def\undertext#1{\vtop{\hbox{#1}\kern 1pt \hrule}}

\def\VEV#1{\left\langle\,#1\,\right\rangle}
\def\tr{\hbox{tr}\,}

\def\pbyp#1#2{\frac{\partial#1}{\partial#2}}

\def\be{\begin{equation}}
\def\ee{\end{equation}}
\def\bea{\begin{eqnarray} & &}
\def\eea{\end{eqnarray}}

\def\rf#1{(\ref{#1})}
\def\rfs#1{Eq.~(\ref{#1})}

\def\ad{a^\dagger}

\def\p{{\bf p}}

\def\a {{\hat a}}
\def\ad {{\hat a^\dagger}}

\def\k {{\bf k}}
\def\q {{\bf q}}

\begin{document}
\title{Quench-induced Floquet topological $p$-wave superfluids} 

\author{Matthew S. Foster$^1$, Victor Gurarie$^2$,  Maxim Dzero$^3$, Emil A. Yuzbashyan$^4$ }
\address{$^1$Department of Physics and Astronomy, Rice University, Houston, Texas 77005, USA}
\address{$^2$Department of Physics, University of Colorado, Boulder, CO 80309, USA}
\address{$^3$Department of Physics, Kent State University, Kent, Ohio 44242, USA}
\address{$^4$Center for Materials Theory, Department of Physics and Astronomy, Rutgers University, Piscataway, New Jersey 08854, USA}

\date{\today}

\begin{abstract}
{
Ultracold atomic gases in two dimensions tuned close to a $p$-wave Feshbach resonance were 
expected to exhibit topological superfluidity, but these were found to be experimentally 
unstable. We show that one can induce a topological Floquet superfluid if weakly interacting 
atoms are brought suddenly close (``quenched'') to such a resonance, in the time before the 
instability kicks in. The resulting superfluid possesses Majorana edge modes, yet differs from 
a conventional Floquet system as it is not driven externally. Instead, the periodic modulation 
is self-generated by the dynamics.
}

\end{abstract}


\maketitle


An intense search is underway to identify experimental realizations of topological 
superconductors, exotic quantum states of matter that carry robust energy currents 
along their boundaries. Topological superconductors can host Majorana fermions, fractional 
particles whose discovery could enable fault-tolerant quantum computation.

A system of identical fermionic atoms confined to two dimensions and 
interacting attractively through a $p$-wave Feshbach resonance  
was predicted to form a ``$p_x + i p_y$'' superfluid \cite{Gurarie2007}. 
Here, $p_x+ip_y$ refers to a particular symmetry of 
the superfluid order parameter; phases of other symmetries are also possible, 
but are not as energetically favorable \cite{Anderson1961,Gurarie2005,Yip2005}.
The excitation spectrum \cite{Read2000} of such a $p_x+ip_y$ state is 
fully-gapped, 
as long as the chemical potential $\mu$ is not zero. 
However, 
if 
the sample possesses a boundary 
and if 
$\mu>0$,
then gapless excitations appear at the superfluid edge \cite{VolovikBook1,Read2000},
propagating in a particular direction.
The edge excitations can be thought of as a one-dimensional band of
Majorana fermions. When the $p_x+ip_y$ superfluid 
is deformed by a perturbation that does not close the bulk energy gap, 
the gapless boundary excitations remain and retain their properties. 
These are said to be ``topologically protected,''
and the phase of the superfluid with $\mu > 0$ 
is a two-dimensional topological superconductor.
This picture applies to the weak coupling BCS regime;
the strong pairing BEC phase 
\cite{Leggett1980,Nozieres1984} 
has $\mu < 0$ and is 
topologically trivial. These are separated
by a quantum critical point at $\mu = 0$ \cite{Read2000,Gurarie2005}.

Several attempts were made to create the $p$-wave superfluid 
experimentally in a gas of fermionic $^{40}$K or $^6$Li atoms. 
Unfortunately these gases were found to be unstable due to losses involving three-body 
processes \cite{Salomon,JL2008,Levinsen2008}, with the lifetime 
$\tl$
ranging from a few milliseconds in 
$^{40}$K 
\cite{Gaebler2007}
to about 20ms in 
$^6$Li \cite{Ticknor2008,Inada2008}
at a particle density corresponding to a Fermi energy of about $10$KHz.
This instability prevents the gas
from reaching its ground state; instead it decays with atoms simply 
leaving the trap where the gas is held. Interestingly, a very \emph{weakly interacting} 
$p$-wave gas is predicted to be significantly more stable \cite{JL2008}, 
yet such a gas would also have a very small gap,
potentially preventing direct observation of its topological properties. 

In this Letter, we show that one can induce a topological Floquet superfluid \cite{Lindner2011,Kitagawa2011}  
if weakly interacting atoms are brought suddenly close to a Feshbach resonance,
in the time before the instability kicks in. 
We build off of our recent work \cite{PwaveLong}, in which we determined
the exact asymptotic dynamics of a BCS p-wave superfluid following a 
quantum quench.
Specifically, we propose to start with a weakly interacting gas of $^{40}$K or $^6$Li 
and then suddenly tune the interaction strength to the desired value by means 
of a Feshbach resonance. The gas would evolve in an out-of-equilibrium fashion
from its initial state.
Our results determine the evolution over the time scale before the instability 
destroys the gas. Note that the ratio of the lifetime 
$\tl$
to the inverse Fermi energy 
$t_F \equiv 2 m / p_F^2$
can be as high as 
$200$, which gives plenty of room for the gas to evolve and reach a 
quasi-stationary state before 
decaying, as will be elaborated below. 
The types of superfluid states that we describe have topologically-trivial s-wave 
analogs \cite{Barankov04,galperin,WarnerLeggett05,YuzbashyanAltshuler05a,YuzbashyanAltshuler05b,YuzbashyanAltshuler06,BarankovLevitov06,DzeroYuzbashyan06}.
An exciting recent development is the observation of non-equilibrium order parameter dynamics 
in superconducting thin films \cite{Shimano13}. 

\begin{figure}[b!]
\centering
\includegraphics[width=0.4\textwidth]{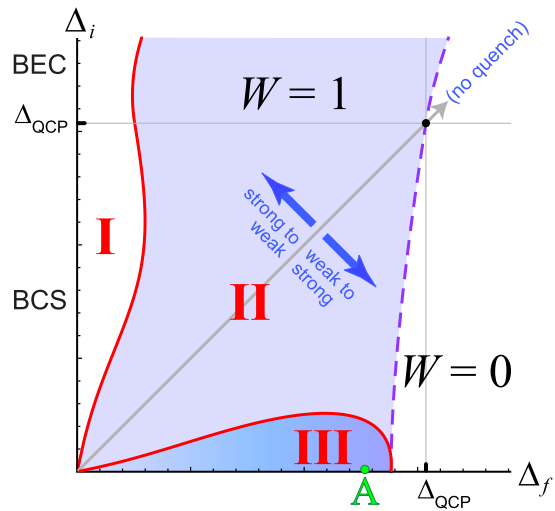}
\caption{
	Phase diagram showing the three regimes ({\bf I}--{\bf III})
	of non-equilibrium superfluidity reached after a quantum quench in a $p$-wave gas \cite{PwaveLong}. 
	Each point in this phase diagram represents a particular quench, wherein 
	one 
	takes an initial state with order parameter amplitude $\Delta_i$, and subsequently 
	ramps the strength of attractive atom-atom interactions to weaker or stronger pairing. 
	The initial state is specified via the vertical axis.
	The horizontal axis measures $\Delta_f$, which is the amplitude one would find in the 
	\emph{ground state} of the post-quench Hamiltonian.
	The diagonal line $\Delta_i = \Delta_f$ is the case of no quench;
	$\Dqcp$ locates the BCS-BEC ground state transition \cite{SUPINFO}.
	Each off-diagonal point to the left (right) of this line denotes a
	particular quench from stronger-to-weaker (weaker-to-stronger) pairing. 
	The Regions labeled {\bf I}, {\bf II}, {\bf III} denote three different regimes of 
	non-equilibrium superfluid dynamics. 
	For a strong-to-weak quench in {\bf I}, the order parameter $\Delta(t)$ decays to zero.
	A quench in {\bf II} leads to a non-zero steady-state order parameter amplitude. 
	A weak-to-strong quench in {\bf III} induces persistent oscillations in $|\Delta(t)|$. 
	$W$ denotes the winding number described in the text.
	Quenches in {\bf II} with $W = 1$ and in {\bf III} produce topological states. 	
	Point ``A'' specifies a quench from very weak initial pairing that produces a Floquet topological 
	state, which could be accessible experimentally.
}
\label{fig1}
\end{figure}

Depending on the initial state and the strength of the quench, the resulting out-of-equilibrium 
superfluid may find itself in one of three regimes \cite{PwaveLong}: 
a steady state with a vanishing 
order parameter $\Delta(t) \rightarrow 0$ as $t \rightarrow \infty$ (Region {\bf I}), 
a state with $\Delta(t) \rightarrow \Delta_\infty$, a non-zero constant (Region {\bf II}), and 
a quasi-steady state with an oscillating $\Delta(t)$ (Region {\bf III}). 
The phase diagram of all possible quenches of the superfluid is shown in 
Fig.~\ref{fig1}. 

To realize a topological superfluid in an ultracold gas, 
the most relevant quenches are those in Region {\bf III}.
An initial state with weak pairing is prepared far from the 
Feshbach resonance (i.e.\ $\Delta_i$ in Fig.~\ref{fig1} is close to zero), where 
three-body losses can be neglected \cite{JL2008}. 
Then the coupling in the Hamiltonian is
quenched close to the resonance, and the system evolves coherently.
Here we show that Region {\bf III} is topologically non-trivial,
and the oscillating order parameter induces Majorana edge modes. 
Region {\bf III} therefore realizes a Floquet topological superfluid,
yet this differs from a conventional Floquet system \cite{Lindner2011,Kitagawa2011,Gu2011,Levin2012}
as it is not driven externally. 
Instead, the periodic modulation is self-generated by the dynamics.

We briefly recount the setup of the problem from Ref.~\cite{PwaveLong}. 
Neglecting the terms responsible for the losses, the 
gas can be described by the Hamiltonian \cite{Gurarie2007}
\be \label{eq:ham} 
	\hat H = \sum_{\p} \frac{p^2}{2m} \ad_\p \a_\p 
	- \frac{\lambda}{V} \sum_{\p,\q, \k} \q \cdot \k  \, \ad_{\frac \p 2 + \q} \ad_{\frac \p 2 - \q} \a_{\frac \p 2 - \k} \a_{\frac \p 2 + \k}.
\ee
Here $\ad_\p$ and $\a_\p$ create and annihilate fermions of mass $m$ with momentum $\p$, 
$\lambda > 0$ denotes their interaction strength, and $V$ is the volume of the system. 
In the following, we
imagine fixing the coupling strength to some initial value $\lambda=\lambda_i$,
and preparing the system of atoms in the corresponding ground state. Then we suddenly change (quench) the 
coupling to a different value $\lambda_f$. We 
then evaluate
how the state of the fermions evolves in time after this quench. 

We compute the dynamics of Eq.~(\ref{eq:ham}) within self-consistent
BCS mean field theory, as governed by the Hamiltonian
\be 
	\label{eq:eff} 
	\hat H_{\rm eff} = \sum_{\p} \frac{p^2}{2m} \ad_\p \a_\p
	+  
	\left[
	\Delta(t) {\sum}^\prime_\p p \, \ad_\p \ad_{-\p} + {\rm h.c.} 
	\right].
\ee
Here the symbol ${\sum}^\prime$ signifies that the summation is restricted to $\p$ that satisfy $p_x>0$,
and $\Delta(t)$ is the amplitude of the gap function, defined as
\be \label{eq:deltaeff} 
	\Delta(t) = 
	-
	\frac{2 \lambda}{V} {\sum}^\prime_\p \, p \,  \VEV{\a_{-\p} \a_{\p}}.
\ee
The time-dependent state of the fermions is of the BCS form \cite{Barankov04}
\be \label{eq:statet}   
	\left| \Omega(t) \right> = {\prod_\p}^\prime \left[ u_p(t) + v_p(t)\,  \ad_{\p} \ad_{-\p} \right] \left| 0 \right>,
\ee 
where $\left|0 \right>$ is the vacuum, and 
Eqs.~\rf{eq:eff} and \rf{eq:deltaeff} reduce to nonlinear 
differential equations satisfied by $u_p(t)$ and $v_p(t)$. 
We solve these equations exactly, exploiting the integrability of
the equations of motion \cite{YuzbashyanAltshuler05a,YuzbashyanAltshuler05b,YuzbashyanAltshuler06,BarankovLevitov06,Sierra2009}.
The solution employs a Lax spectral method. The analysis closely parallels 
work done for the corresponding $s$-wave problem in three dimensional space 
\cite{Barankov04,galperin,WarnerLeggett05,YuzbashyanAltshuler05a,YuzbashyanAltshuler05b,YuzbashyanAltshuler06,BarankovLevitov06,DzeroYuzbashyan06}. 
This was carried out in Ref.~\cite{PwaveLong} and the results are shown in Fig.~\ref{fig1}.

The mean field approach differs from \rfs{eq:ham} in three ways. 
First, the interaction terms in \rfs{eq:ham} with $\p \not = 0$ have been removed. 
This is a standard approximation in the theory of superconductivity: 
the only terms retained in the interaction are those responsible for the 
pairing of the fermions into Cooper pairs or strongly-bound 
molecules which then Bose condense.
Since our goal is to predict dynamics from a given initial state,
our results will hold over a time interval in which the effects of 
neglected terms remain small. This is the minimum of $\tl$ and $\tpb$, 
where $\tpb$ is the pair-breaking lifetime induced by $\p \not = 0 $ terms
\cite{Barankov04,VolkovKogan74}. 

Second, the mean field approach neglects quantum fluctuations in $\Delta(t)$.
Without pair-breaking terms, this is exact in the thermodynamic limit. 
It is well known \cite{Richardson64,Dukelsky04} that fluctuations induce only 
finite-size corrections. The reason is that $\Delta(t)$ is a \emph{global}, 
not merely a local mean field. It becomes macroscopic and classical if the number of 
fermions is sufficiently large and the system exhibits superconducting order 
(irrespective of equilibrium).
Finally, we assume an initial state with $p_x+ip_y$ symmetry. In mean field theory 
this ``projects out'' the $p_x - i p_y$ channel, so that it does not participate in the dynamics. 
The change of variables $\a_{- \k} \a_{\k} \rightarrow e^{i \phi_{\k}} \a_{- \k} \a_{\k}$ 
leads to Eqs.~(\ref{eq:eff})--(\ref{eq:statet}) \cite{PwaveLong};
$\phi_{\k}$ is the polar angle.

Let us now examine Region {\bf III}, of particular interest here. 
In this 
case, the order parameter asymptotes to
\be \label{eq:region2} 
	\Delta(t) = \Delta_\infty(t) \, e^{-2 i \mu_{\infty} t}, 
\ee 
where $\Delta_\infty(t)$ is a complex-valued periodic
function of time with some period $T$, and $\mu_{\infty}$ is
a real constant. These are completely determined by the particular 
quench specified by $\{\Delta_i,\Delta_f\}$ \cite{PwaveLong,SUPINFO}. 
In general, $T$ and $\pi/\mu_\infty$ are incommensurate periods.
By absorbing $\mu_\infty$ into the phase of the operators 
$\a_\p$ and $\ad_\p$, we can map our effective
Hamiltonian in \rfs{eq:eff} to a
superconductor with an 
oscillatory complex-valued order parameter 
$\Delta_\infty (t)$ and chemical potential $\mu_\infty$. 
A useful quantity to characterize such a superconductor is its retarded 
Green's function ${\cal G}(t,t')$, defined as the solution of the matrix 
Bogoliubov-de Gennes equation \cite{SUPINFO}
\be \label{eq:BdG}  
	i \pbyp{{\cal G}}{t} - {\cal H} {{\cal G}}  
	= 
	\delta(t-t'), 
	\  
	{\cal H } 
	= 
	\left( \begin{matrix} { \frac{p^2}{2m} - \mu_\infty & \Delta_\infty \, p \, e^{i \phi_{\bf p}} \cr \Delta^*_\infty \, p\, e^{-i \phi_{\bf p}} &- \frac{p^2}{2m}+\mu_\infty } \end{matrix} \right),
\ee where $\phi_{\bf p}$ is the angle ${\bf p}$ forms with the positive $x$-direction. 
This equation is identical to 
that for a driven
superconductor with a gap function imposed to be a given function of time. 
We 
must still
keep in mind that we are describing a strongly 
out-of-equilibrium 
state, with $\Delta_\infty$ determined by
the contributions of many fluctuating Cooper pair amplitudes such as $\VEV{\a_{-\p} \a_{\p}}(t)$ 
in \rfs{eq:deltaeff}. 

Interestingly, in Region {\bf II} where $\Delta_{\infty}$ is a constant, the corresponding BdG 
equations formally match those of an equilibrium superconductor. The equilibrium superconductor 
is characterized by a topological number $W$, which depends solely on the sign of the chemical 
potential \cite{Read2000}. If the chemical potential is positive, $W=1$ and the system, while gapful 
in the bulk, is known to have gapless edge states. 
Ref.~\cite{Essin2011} argued that any retarded Green's function with a topological number $W=1$ 
when computed in a geometry with a boundary will have poles corresponding to gapless excitations 
in the boundary. Therefore, even the far from equilibrium superconductor 
discussed here will have topologically protected edge states as long as $\mu_{\infty}$ is positive \cite{SUPINFO}.
The range of positive $\mu_{\infty}$ is shown in Fig.~\ref{fig1} as a subregion of Region {\bf II} with $W=1$. 

Unitary time evolution is a smooth rotation of the initial state. It may therefore appear surprising that 
a quench can induce a change in a winding number within a finite time interval. In fact, one must distinguish 
two different notions of topology here. The topology of the state (pseudospin winding) does not change, 
but that of the effective single particle Hamiltonian can
($W$, as defined via the retarded Green's function in 
\cite{SUPINFO}). 
The
Green's function
determines the 
frequency spectrum that appears when transitions are driven by an external probe,
while the 
state
encodes the occupation of the modes \cite{PwaveLong}. 

In Region {\bf III}, $\Delta_{\infty}(t)$ is a complex-valued periodic 
function of time that can be determined analytically 
\cite{Barankov04,YuzbashyanAltshuler05a,YuzbashyanAltshuler05b}.
The parameters of this function including its turning points and
the period $T$ are computed for a particular quench by solving a 
certain transcendental equation \cite{PwaveLong,SUPINFO}.
A periodically driven 
system can be topological in the Floquet sense, as was recently discussed in the 
literature \cite{Lindner2011,Kitagawa2011,Gu2011}. What this implies is that one needs to 
construct its Green's function $U(T)={\cal G}(t+T,t)$ with $t$ being arbitrary 
[but sufficiently large so that the large time asymptotic for 
$\Delta(t)$
applies]. The edge states of this system are then the eigenstates of $U(T)$, with their 
energy related to the eigenvalues of $U(T)$. More precisely, the eigenvalues of $U(T)$, 
a unitary operator, 
assume the
form 
of $\exp(-i \epsilon T)$, where $\epsilon$ 
is such an
energy 
level,
taken to reside
in the compact interval $[-\pi/T, \pi/T]$. 
These 
``quasi-energies'' are similar to the crystalline quasi-momentum 
in systems periodic in space (while here the Hamiltonian is periodic in time). 

It is possible to extract whether this system is topological by 
analyzing \cite{Levin2012} 
the winding of ${\cal G}(t,t')$. In practice, this may not be easy to do. 
Instead, given the analytic expression for the 
time-dependent 
$\Delta_{\infty}(t)$ 
associated to a particular quench \cite{PwaveLong},
we solved the Bogoliubov-de-Gennes equation numerically in the cylinder geometry 
(periodic in one direction, with a hard wall boundary in the other) \cite{SUPINFO}. 
After computing
$U(T)$ in this geometry,
we extracted its eigenvalues and checked whether the edge states appear. 
By doing this at various points in Region {\bf III}, we can map out
the topological character of this dynamical phase.
At the boundary of Regions {\bf III} and {\bf II} 
where $\Delta_{\infty}$ becomes a 
constant, we know
from Fig.~\ref{fig1} that the system is topological. 
Thus we expect that within the Region {\bf III} close to the boundary with Region {\bf II},
the topological aspects of the phase (the boundary states) remain, even though the winding number $W$ may change, 
as was recently pointed out \cite{Levin2012}, while deep within Region {\bf III} there 
might in principle be non-topological domains or domains with a different 
topology from that in {\bf II}. 

\begin{figure}
\centering
\includegraphics[width=0.43\textwidth]{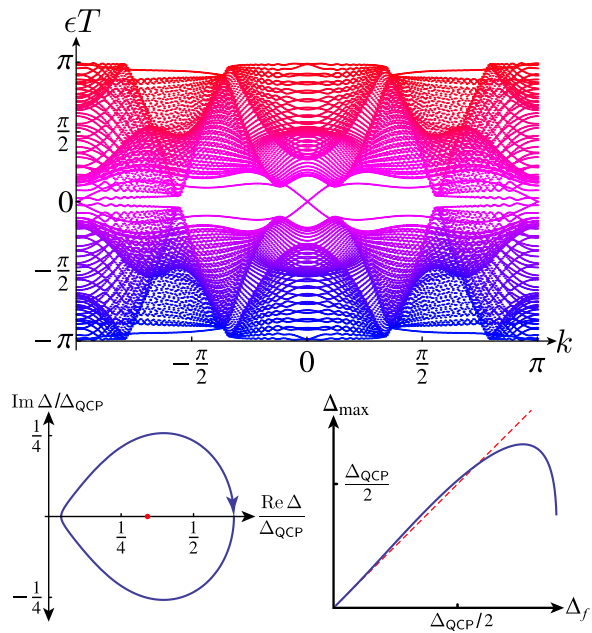}
\caption{
	Majorana edge modes for a quench-induced time-dependent state of $p$-wave superfluidity.
	The Floquet spectrum (top) of $\ln U(T)$ for a system on a finite cylinder 
	is plotted in the large time asymptotic regime for a quench in Region {\bf III}, 
	point ``A'' in Fig.~\ref{fig1}. The horizontal axis represents the momentum along 
	the boundary, the vertical axis the quasienergies multiplied by the period of oscillations, 
	both ranging from $-\pi$ to $\pi$. The edge states can be clearly seen crossing in the 
	center of the figure. The bottom left shows the orbit swept by  
	$\Delta_{\infty} (t)$ in the complex $\Delta$ plane at this point in Region {\bf III}.
	The bottom right plots the 
	orbital
	maximum 
	$\Delta_{\mathrm{max}} \equiv \max|\Delta(t)|$ 
	as a function of $\Delta_f$
	for quenches from very weak initial pairing ($\Delta_i \rightarrow 0$) in {\bf III}.
	The dashed line is $\Delta_{\mathrm{max}} = \Delta_f$.
	}
\label{fig2}
\end{figure}

We find that edge states are present no matter where within Region 
{\bf III} we look.
Fig.~\ref{fig2} exhibits
a typical spectrum for $U(T)$ 
at a point deep within Region 
{\bf III},
indicated in Fig.~\ref{fig1} as 
point A. This 
quench is located at
$\Delta_i/\Delta_{\rm QCP}= 0.0065$, 
$\Delta_f/\Delta_{\rm QCP}=0.83$. 
To generate the plot shown in 
Fig~\ref{fig2}, 
we placed the superfluid on the lattice, with 50 lattice constants 
within the width of the cylinder \cite{SUPINFO}. 
The hopping amplitude on this lattice was chosen to be 
$J = 1/2$,  so that the system would be below half 
filling, yet the total bandwidth 
was as small as possible to prevent the spectrum from folding too many 
times onto itself and obscuring the graph. 
Crossing edge states in the center of the figure prove that the time-dependent superfluid for this 
particular quench is topological in the Floquet sense. We conjecture that the entire Region {\bf III} 
is topological, but proving this  requires further work.

A natural quench from the point of view of experiment would 
start
from the non-interacting Fermi gas at $\Delta_i=0$. Such a quench is much harder to describe than 
those
studied here so far. Technically, the zero-temperature Fermi-Dirac distribution is a point of unstable 
equilibrium for the classical equations of motion studied above, so naively it 
does not evolve in time.
In reality, quantum or thermal fluctuations will generate 
an
initial condition with nonzero 
$p_x + i p_y$ and $p_x - i p_y$ order parameter amplitudes, and
these will compete in the subsequent dynamics.
The precise outcome 
is difficult to predict. 
Instead, we assume that it is 
possible to first switch on very weak attraction which results in some initial very small yet 
nonzero $\Delta_i$ 
of pure $p_x + i p_y$ type.
Then we
quench this state into a far stronger interacting 
regime; 
as long as the quench resides within Region {\bf III} we expect the resulting state to be in the 
topological Floquet phase. (Point A where $\Delta_i \ll \Delta_f$ is a good example of 
such a quench.) At the same time, if the interactions after the quench are stronger than the 
threshold depicted Fig.~\ref{fig1}, we will end up in the non-topological ($W = 0$) 
domain of Region {\bf II}. 

We conclude with a discussion of relevant time scales.
The main effect of the $p \neq 0$ terms in \rfs{eq:ham} is to mediate pair-breaking collisions \cite{Barankov04,VolkovKogan74},
associated to a 
rate $1/\tpb$.
For quenches confined to the BCS regime, the lifetime can be estimated using Fermi liquid theory \cite{VolkovKogan74,Zwierlein05,BarankovLevitov06},
leading to
$
	\tpb / t_F
	\sim
	\left[\e_F/\Emin(\Delta_f)\right]^2,	
$
where $t_F = 1/\e_F$ is the inverse Fermi energy and 
$\Emin(\Delta) = \Delta \sqrt{2 \mu - \Delta^2}$
is the ground state 
quasiparticle energy gap. Quenches in Region {\bf III} that produce topological Floquet
states reside entirely within the BCS regime; for these,
the ratio $\tpb/t_F$ can easily be an order of magnitude,
and grows rapidly larger as $\Delta_f$ is reduced. 
The inverse 
three-body loss rate 
can be estimated to be
$
	t_3 / t_F \sim (\ell / b)^\alpha, 
$
where $\ell \sim 1000$ nm ($b \sim 5$ nm) is the interparticle separation (Van der Waals length) \cite{JL2008,Levinsen2008}. 
Near resonance, the exponent $\alpha$ = 1, with $t_3 \sim 20$ ms in experiments \cite{Ticknor2008,Inada2008}.
However, towards the weak BCS regime $t_3$ becomes orders of magnitude larger with $\alpha = 4$ \cite{JL2008},
making three-body losses essentially irrelevant for the creation of a weakly paired initial state. 

In numerical simulations of our model \cite{PwaveLong}, we find that the asymptotic 
behavior is reached very quickly in Region {\bf III} over a time $t \lesssim t_F$. 
For quenches from weak initial pairing
with $\Delta_i \ll \Delta_f$,
the period $T$ of oscillations in the
order parameter magnitude can be estimated as \cite{PwaveLong,SUPINFO} 
\be
	T \sim \frac{2}{\Emin(\Delta_f)} \ln \left[\frac{4 \e_F}{\Lambda} \frac{\Emin(\Delta_f)}{\Emin(\Delta_i)} \right] 
	\sim \sqrt{t_{F}\tpb},
\ee
where $\Lambda$ is an ultraviolet energy cutoff.	
In the BCS regime we always have $\e_F > 2 \Emin$, so that
\be
		t_{F} < T < \min(\tpb,t_3),
\ee	
where $t_3$ is associated to the (larger) post-quench coupling strength.
This is the window in which the
topological non-equilibrium steady state can be realized. Decreasing the final pairing 
strength 
[increasing $\e_F/\Emin(\Delta_f)$]
may increase the relative size of the window,
at the cost of decreasing the detectable maximum of $|\Delta(t)|$
(see Fig.~\ref{fig2}).

Quench-induced 
topological edge states could be detected using RF spectroscopy type experiments. 
An open question is 
whether these
types of topological steady states support 
the kind of quantized thermal conductance expected in an 
equilibrium topological $p$-wave superconductor. 
Calculating energy transport and exploring possible quantized 
out-of-equilibrium transport phenomena remains the subject for future work.

V.G.\ thanks Y.\ Castin for useful discussions.	
This work was supported in part by the NSF under Grants No.~DMR-0547769 (M.S.F. and E.A.Y.),
DMR-1205303 and PHY-1211914 (V.G.), 
the NSF I2CAM International Materials Institute Award, Grant No.~DMR-0844115 (M. D.),
by the David and Lucile Packard Foundation (M.S.F. and E.A.Y.), and by the Welch Foundation 
under Grant No.~C-1809 (M.S.F.).

\vskip 1cm
{\center {\centerline { \bf SUPPLEMENTAL MATERIAL}}}
\vskip .3cm


\section{Scales and units}

We introduce the interaction parameter $g \equiv \lambda m /2$, which has units of length-squared.
After dropping pair-breaking ($\vex{p} \neq 0$) terms, the Hamiltonian in Eq.~(1) becomes
\begin{eqnarray}
	\hat H 
	&=& 
	\frac{1}{m}
	\bigg\{
	\sum_{\p} \frac{p^2}{2} \ad_\p \a_\p 
	\nonumber\\
	&&
	- 
	\frac{g}{V} 
	{\sum_{\q, \k}} \left[(q^x + i q^y)(k^x - i k^y) + \textrm{c.c.}\right] \, \ad_{ \q} \ad_{ - \q} \a_{ - \k} \a_{\k}
	\bigg\},
	\nonumber\\
\end{eqnarray}
where $\textrm{c.c.}$ denotes the complex conjugate. 
We set $m = 1$, since it merely specifies the units for the Hamiltonian.
For a $p_x + i p_y$ superfluid with order parameter amplitude $\Delta$, 
the energy of a bulk quasiparticle excitation with momentum $\vex{p}$ is
\be
	E_p = \sqrt{(p^2/2 - \mu)^2 + \Delta^2 p^2},
\ee
where $\mu$ is the chemical potential. $\Delta$ carries inverse length units.

The BCS mean field equations relating the ground state $\Delta$ and $\mu$ to the 
coupling strength $g$ and fixed particle density $n$ are given by \cite{SM-PwaveLong}
\be\label{MFG}
	\frac{1}{g}
	=
	\frac{1}{g_{\msf{QCP}}}
	+
	\frac{\mu}{\pi} \ln\left(\frac{2 \Lambda e}{\Delta^2}\right),
	\quad
	\mu 
	=
	2 \pi n - \frac{\Delta^2}{2} \ln\left( \frac{2 \Lambda}{e \Delta^2} \right),
\ee
where $g_{\msf{QCP}}$ denotes the coupling strength at the BCS-BEC transition 
$\mu = 0$, 
\[
	\frac{1}{g_{\msf{QCP}}} 
	= \frac{\Lambda}{\pi} - 4 n.
\]
Eq.~(\ref{MFG}) applies to the BCS side with $\mu \geq 0$; 
$\Lambda$ is an ultraviolet energy cutoff. 
These equations are valid up to corrections of order $1/\Lambda$.

The natural scale for $\Delta$ is the value at the BCS-BEC quantum phase transition \cite{SM-PwaveLong},
\be
	\Dqcp 
	\simeq
	\sqrt{
	\frac{4 \pi n
	}{
	\ln\left(\frac{\Lambda}{2 \pi n e} \right)  + \ln\left[\ln\left(\frac{\Lambda}{2 \pi n e}\right)\right]
	}	
	}.
\ee


\section{Retarded Green's function}

The matrix retarded Green's function $\mathcal{G}(t,t')$ in Eq.~(6) is defined as follows:
\begin{eqnarray}\label{GDef}
	\mathcal{G}(t,t')
	&\equiv&
	-i
	\left[
	\begin{array}{cc}
	\langle \{\hat{a}_{\vex{p}}(t),\hat{a}_{\vex{p}}^\dagger(t')\} \rangle 
	&
	\langle \{\hat{a}_{\vex{p}}(t),\hat{a}_{-\vex{p}}(t')\} \rangle 
 	\\
	\langle \{\hat{a}^\dagger_{-\vex{p}}(t), \hat{a}_{\vex{p}}^\dagger(t')\} \rangle 
	& 
	\langle \{\hat{a}^\dagger_{-\vex{p}}(t),\hat{a}_{-\vex{p}}(t')\} \rangle 
	\end{array}
	\right]
	\nonumber\\
	&&
	\phantom{-i}
	\times
	\theta(t - t'),
\end{eqnarray}
where $\{A,B\} \equiv A B + B A$ is the anticommutator.

Following a quench at time $t = 0$, when $\Delta(t) \rightarrow \Delta_\infty$ (constant, Region {\bf II} in Fig.~1 
of the main text), one has $\mathcal{G}(t,t') = \mathcal{G}(t - t')$ for sufficiently large 
times $t$ and $t'$. Then Eq.~(6) is identical to its equilibrium counterpart. 
This is true despite the fact that the post-quench coupling constant $\lambda_f$ is by no means 
equal to that needed to produce $\Delta_\infty$ and $\mu_\infty$ in equilibrium \cite{SM-PwaveLong}.
To determine the presence or absence of Majorana edge modes in the strip geometry, we
compute the bulk winding number \cite{SM-Volovik88,SM-Gurarie11,SM-EssinGurarie11}
\begin{eqnarray}
	\label{W}
	W
	&\equiv&
	\frac{\epsilon_{\alpha \beta \gamma}}{3!}
	\int_{-\infty}^{\infty}
	d \omega
	\int 
	\frac{d^2 \vex{p}}{(2 \pi)^2}
	\nonumber\\
	&&
	\times
	\tr\left[
	\G^{-1} \left(\partial_\alpha \G\right)
	\,
	\G^{-1} \left(\partial_\beta \G\right)
	\,
	\G^{-1} \left(\partial_\gamma \G\right)
	\right],
\end{eqnarray}
where $\G \equiv \G_{\vex{p}}(i \omega)$ is the Fourier transform of $\mathcal{G}_{\vex{p}}(t - t')$,
continued to imaginary frequency. Edge modes are signaled by $W$ = 1, while quenches that
produce trivial states have $W = 0$. These are shown in Fig.~1.


\section{Region {\bf III} dynamics}

The exact solution in \cite{SM-PwaveLong} gives the asymptotic $\Delta(t)$ 
in Region {\bf III} in terms of amplitude and phase variables,
\[
	\Delta(t) \equiv \sqrt{\Rhot} \exp(- i \phi ).
\]
These satisfy
\begin{eqnarray}
	\dot{\Rhot}^2
	&=&
	(\Rhot_{+} - \Rhot)(\Rhot - \Rhot_{-})
	(\Rhot + \wRhot_{+})
	(\Rhot + \wRhot_{-}),
	\nonumber\\
	\dot{\phi} 
	&=&
	\frac{3}{2}
	\Rhot
	+
	2 \msf{m}
	-	
	\frac{\psi}{\Rhot}.
\end{eqnarray}
The parameters in these equations can be expressed as
\begin{eqnarray}\label{RhotParams}
	\Rhot_{\pm}
	&\equiv&
	\frac{1}{2} 
	\left[
	\sqrt{\left(|u_1| - u_{1,\rr}\right)} \,
	\pm 
	\sqrt{\left(|u_2| - u_{2,\rr}\right)} \,
	\right]^2,
	\\
	\wRhot_{\pm} 
	&\equiv&
	\frac{1}{2} 
	\left[
	\sqrt{\left(|u_1| + u_{1,\rr}\right)} \,
	\pm 
	\sqrt{\left(|u_2| + u_{2,\rr}\right)} \,
	\right]^2,
	\\
	\mathsf{m} 
	&=&
	\frac{1}{4}
	\left(u_{1,\rr} + u_{2,\rr}\right),
	\\
	\psi
	&=&
	\frac{1}{8}
	\left[
	-
	(u_{1,\ii}^2 + u_{2,\ii}^2) 
	-
	2 u_{1,\rr} u_{2,\rr}
	+
	2
	|u_{1}||u_{2}|
	\right]\!,
\end{eqnarray}
where $|u_a| = \sqrt{u_{a,\rr}^2 + u_{a,\ii}^2}$.
The four parameters $u_{\{1,2\},\rr}$ and $u_{\{1,2\},\ii}$ locate the two isolated
pairs of roots of the spectral polynomial for a quench in Region {\bf III} 
\cite{SM-PwaveLong}. These are uniquely determined by solving a certain
transcendental equation for a particular quench specified by 
$(\Delta_i,\Delta_f)$ in {\bf III}.

The period of $\Delta(t)$ is given by \cite{SM-PwaveLong}
\begin{eqnarray}\label{period}
	T 
	&=&
	\frac{2}{\alpha}
	K\left( 
	\frac{u_{1,\ii} u_{2,\ii}}{\alpha^2} 
	\right),
	\nonumber\\
	\alpha
	&=&
	\frac{1}{2}
	\sqrt{(u_{1,\rr} - u_{2,\rr})^2 + (u_{1,\ii} + u_{2,\ii})^2},
\end{eqnarray}
where $K(M)$ is the complete elliptic integral of the first kind (and $M = k^2$).


\section{Lattice regularization for Floquet}

To obtain the Floquet spectrum in Fig.~2 for the quench labeled ``A'' in Fig.~1, 
we employed the following procedure. 
First, we computed the parameters in Eq.~(\ref{RhotParams}) from the isolated roots
for the quench \cite{SM-PwaveLong}. This determines the function $\Delta_{\infty}(t)$ and
the real constant $\mu_{\infty}$ in Eq.~(5).
The orbit of the resulting $\Delta_{\infty}(t)$ is plotted in the lower left of Fig.~2. 
Here we have used the high-energy cutoff $\Lambda = 100 \pi n$ to fix $\Dqcp$. 

We then diagonalized the Floquet operator $U(T)$ using the following lattice regularization of
the $p_x + i p_y$ Bogoliubov-de Gennes Hamiltonian:
\begin{eqnarray}
	H
	&=&
	\sum_{m = 1}^{M}
	\sum_{n = -\infty}^{\infty}
	\bigg[
	-
	J
	\,
	c^\dagger_{m,n}
	\left(\!
	\begin{array}{l}
	c_{m+1,n}
	+
	c_{m-1,n}
	\\
	\;
	+
	c_{m,n+1}
	+
	c_{m,n-1}
	\end{array}
	\!\right)
	\nonumber\\
	&&
	+
	(4 J - \mu_\infty)
	c^\dagger_{m,n}
	c_{m,n}
	\bigg]
	\nonumber\\
	&&
	+
	\frac{i \Delta_{\infty}(t)}{4}	
	\sum_{m = 1}^{M}
	\sum_{n = -\infty}^{\infty}
	c_{m,n}
	\left(
	\begin{array}{l}
	c_{m+1,n}
	-
	c_{m-1,n}
	-
	\\
	\;
	i
	c_{m,n+1}
	+
	i
	c_{m,n-1}
	\end{array}
	\right)
	+ 
	\textrm{h.c.}
	\nonumber\\
\end{eqnarray}
Here h.c.\ denotes the Hermitian conjugate.

\end{document}